\documentclass[prb,twocolumn,showpacs,preprintnumbers,amsmath,amssymb,superscriptaddress]{revtex4-1}

\usepackage{amsmath,amssymb,amsfonts,mathrsfs} 	% Typical maths resource package
\usepackage{graphicx}
\usepackage{subfigure}

\renewcommand{\[}{\begin{equation}}
\renewcommand{\]}{\end{equation}}
\def\bea{\begin{eqnarray}}
\def\eea{\end{eqnarray}}
\def\nn{\nonumber\\}

\newcommand{\emi}[1]{{\rm e}^{-i #1}}
\newcommand{\ei}[1]{{\rm e}^{i #1}}

\newcommand{\intk}{\int_{\rm BZ} \!\!\!\!\!  [d {\bf k}] }
\newcommand{\E}{{\bf E}}

\renewcommand{\j}{{\bf j}}
\newcommand{\p}{{\bf p}}

\renewcommand{\k}{{\bf k}}

\renewcommand{\v}{{\bf v}}
\renewcommand{\r}{{\bf r}}

\newcommand{\vc}{V_{\rm cell}}

\newcommand{\equ}[1]{Eq.~(\ref{#1})}
\newcommand{\eqs}[2]{Eqs.~(\ref{#1}) and (\ref{#2})}

\def\ket#1{\vert#1\rangle}

\def\me#1#2#3{\langle#1| \, #2 \, |#3\rangle}

\def\runtime{(\the\time)\qquad\the\month/\the\day/\the\year}% get current time
\def\today
 {\count10=\year\advance\count10 by -2000 \number\day--\ifcase
  \month \or Jan\or Feb\or Mar\or Apr\or May\or Jun\or
             Jul\or Aug\or Sep\or Oct\or Nov\or Dec\fi--\number\count10}

\def\hour{\count10=\time\count11=\count10
\divide\count10 by 60 \count12=\count10
\multiply\count12 by 60 \advance\count11 by -\count12\count12=0
\number\count10 :\ifnum\count11 < 10 \number\count12\fi\number\count11}

\begin{document}

%%%%%%%%%%%%%%%%%%%%%%%%%%%%%%%%%%%%%%%%%
\title{Drude weight in presence of nonlocal potentials}
%\email{resta[at]iom.cnr.it}

\author{Raffaele Resta}

\affiliation{Istituto Officina dei Materiali IOM-CNR, Strada Costiera 11, 314151 Trieste, Italy}

\affiliation{Donostia International Physics Center, 20018 San Sebasti{\'a}n, Spain}
%\date{\today}

\date{\today}

\begin{abstract}
The nonlocal potential contributes an extra term to the velocity operator; I show here that such term affects the formal expression  of the Drude weight in a nontrivial way. Notably, the present main result fixes a disturbing discrepancy in the Dreyer-Coh-Stengel sum rule [Phys. Rev. Lett. {\bf 128}, 095901 (2022)].

\end{abstract}

\date{run through \LaTeX\ on \today\ at \hour}

%\pacs{xxx}

\maketitle \bigskip\bigskip

%\section{Introduction}

Since the early times of norm-conserving pseudopotentials in the late 1970s it became clear that in presence of nonlocality the electronic velocity operator $\hat\v$ differs from $\hat\p/m$ by an extra term, owing to the basic definition $\hat\v=i[\hat{H},\r]/\hbar$. 

Within band-structure theory one writes the Schr\"odinger Eq. as \[ \hat{H}_\k \ket{u_{j\k}} = \epsilon_{j\k}  \ket{u_{j\k}} , \] where the eigenstates  $\ket{u_{j\k}}$ are lattice-periodical with band energies $\epsilon_{j\k}$, and
 \[ \hat{H}_\k = \emi{\k \cdot \r} \hat{H} \ei{\k \cdot \r} = \frac{1}{2m} \left( \hat{\bf p} + \hbar \k  \right)^2 + \hat{V}_{\k} . \] 
The velocity of a Bloch electron in band $j$ is then \[ \v_{j\k} = \frac{1}{\hbar}\frac{\partial \epsilon_{j\k}}{\partial \k} =  \me{u_{j\k}}{\left(\frac{\hat\p + \hbar \k}{m} + \frac{1}{\hbar} \frac{\partial \hat{V}_{\k}}{\partial \k} \right)}{u_{j\k}} , \label{velocity}
\] and the macroscopic current density in dimension $d$ is (for double band occupancy) \[ \j = - \frac{2e}{\hbar} \sum_j \intk f(\epsilon_{j\k}) \frac{\partial \epsilon_{j\k}}{\partial \k}, \label{j} \] where $[d \k] = d\k/(2\pi)^d$ and $f(\epsilon)$ is the Fermi occupation factor.

Little attention has been devoted so far to nonlocal effects on the Drude weight. The customary expressions---derived in the local case---are\cite{Allen06,rap157}
\[ D^{(\rm local)}_{\alpha\beta} = \frac{2 \pi e^2}{\hbar^2} \sum_{j} \intk  f(\epsilon_{j\k}) \,  \frac{\partial^2 \epsilon_{j\k}}{\partial k_\alpha \partial k_\beta}, \label{band} \]  
\[  D^{(\rm local)}_{\alpha\beta} = - 2 \pi e^2 \sum_j \intk  f'(\epsilon_{j\k}) \; v_{j k_\alpha} v_{j k_\beta} ; \label{Allen} \] the Fermi-volume and the Fermi-surface expressions are related to each other by an integration by parts; \equ{Allen} also admits a semiclassical derivation.\cite{semicl}

To the best of the author's knowledge, \eqs{band}{Allen} are adopted as they stand in the existing literature, even when nonlocal potentials are adopted;\cite{Dreyer22,Marchese23}  in the following, I am going to show that the above expressions  need instead  to be modified in the nonlocal case.

In presence of a macroscopic field $\E$ the vector potential contributes to the velocity by the term $e{\bf A}/c$, linear in time. The time-derivative of \equ{j} is then, to lowest order in $\E$,  \bea \partial_t j_\alpha &=& -\frac{2e}{\hbar}  \sum_j \intk f(\epsilon_{j\k}) \frac{\partial^2 \epsilon_{j\k}}{\partial k_\alpha \partial A_\beta }  \partial_t A_\beta \nn &=& \frac{2ec}{\hbar}  \sum_j \intk f(\epsilon_{j\k}) \frac{\partial^2 \epsilon_{j\k}}{\partial k_\alpha \partial A_\beta }  E_\beta;  \label{acce} \eea this clearly shows that the many-electron system undergoes free acceleration.

In the local case one may exploit  \[ \frac{\partial}{\partial {\bf A}} = \frac{e}{\hbar c} \frac{\partial}{\partial \k} , \label{local} \] in order to obtain  \bea \partial_t j_\alpha &=& -\frac{2e^2}{\hbar^2}  \sum_j \intk f(\epsilon_{j\k}) \frac{\partial^2 \epsilon_{j\k}}{\partial k_\alpha \partial k_\beta }  E_\beta \nn &=& \frac{D^{(\rm local)}_{\alpha\beta}}{\pi}  E_\beta.   \label{rapix} \eea The Fourier transform of \equ{rapix} yields the familiar expression \[  j_\alpha(\omega)=  \sigma^{(\rm Drude)}_{\alpha\beta}(\omega) E_\beta , \] \[ \sigma^{(\rm Drude)}_{\alpha\beta}(\omega) = D^{(\rm local)}_{\alpha\beta} \left[ \delta(\omega) + \frac{i}{\pi \omega}\right] ; \] a derivation can be found e.g. in the Appendix of Ref. \onlinecite{rap160}.

In the nonlocal case \equ{local} no longer holds; it is then expedient to define \bea \tilde\v_{j\k} &=&  \frac{1}{m} \me{u_{j\k}}{\left(\hat\p + \hbar \k \right)}{u_{j\k}} = \frac{c}{e} \me{u_{j\k}}{\frac{\partial \hat{H}_{\k}}{\partial {\bf A}}}{u_{j\k}} \nn &=& \frac{c}{e}  \frac{\partial \epsilon_{j\k}}{\partial {\bf A}} . \label{fv}
\eea
Integrating by parts \equ{acce} one gets \bea \partial_t j_\alpha &=& - \frac{2ec}{\hbar}  \sum_j \intk f'(\epsilon_{j\k}) \frac{\partial \epsilon_{j\k}}{\partial k_\alpha} \frac{\partial \epsilon_{j\k}}{\partial A_\beta } E_\beta; \nn &=&   \frac{D_{\alpha\beta}}{\pi}  E_\beta.\label{acce2} , \eea where the Drude weight is now expressed as \[  D_{\alpha\beta} = - 2 \pi e^2 \sum_j \intk  f'(\epsilon_{j\k}) \; v_{j k_\alpha} \tilde{v}_{j k_\beta}  \label{Allen2} , \] to be compared to \equ{Allen}.

The Drude weight measures the inverse inertia of the many-electron system. In the special case of  a flat potential its value is \[ D^{(\rm free)}_{\alpha\beta} = \frac{\pi e^2 n}{m}\delta_{\alpha\beta},  \] where $n$ is the density of conducting electrons; in the general case one then recasts $D_{\alpha\beta}$ as  \[ D_{\alpha\beta} = \frac{\pi e^2 n^*_{\alpha\beta}}{m}, \label{n*} \] where $n^*_{\alpha\beta}$ has the meaning of the effective electron density contributing to the adiabatic current.\cite{Scalapino93,rap157}

Remarkably, the improved expression of \equ{Allen2} fixes a disturbing discrepancy in the sum rule obeyed by the Born effective charges in metals, recently found by Dreyer, Coh, and Stengel.\cite{Dreyer22} Suppose that the sublattice $s$ is displaced with velocity $\v_s$: the dimensionless Born tensor is by definition (in either insulators or metals) \[ Z^*_{s,\alpha\beta} = Z_s \delta_{\alpha\beta} + \frac{\vc}{e} \frac{\partial j_\alpha}{\partial v_{s\beta}} , \] where $eZ_s$ is the bare core charge and $\vc$ is the unit-cell volume. When the whole lattice is rigidly translated with velocity $\v$, the linearly-induced current density (due to electrons and nuclei) is  \[ j^{(\rm total)}_\alpha = \frac{e}{\vc} \left( \sum_s Z^*_{s,\alpha\beta} \right) v_\beta . \label{zsum} \]  
In the insulating case no current may flow for a rigid translation of the whole solid, whence the basic relationship $\sum_s Z^*_{s,\alpha\beta} =0$, called the acoustic sum rule. In the metallic case some current instead flows; in the reference frame of the nuclei it is carried solely by the electrons, all moving with velocity $-{\bf v}$, and whose effective density is $n^*_{\alpha\beta}$. Therefore  from \equ{n*} \[ j^{(\rm total)}_\alpha = e \,n^*_{\alpha\beta} v_\beta = \frac{m D_{\alpha\beta}}{\pi e} v_\beta  \label{zsum2} .;\] 
by comparing \equ{zsum2} to \equ{zsum} one gets the  Dreyer-Coh-Stengel  sum rule: \[ \frac{1}{\vc} \sum_s Z^*_{s,\alpha\beta} = \frac{m}{\pi e^2}  D_{\alpha\beta} . \]  As said above, the sum rule is exact---even in the nonlocal case---provided \equ{Allen2} is adopted for the Drude weight.

In insulators the Born effective charges manifest themselves in the phonon spectra: they generate a depolarization field $\E$ which contributes to the restoring force at the zone center. No such contribution may exist in metals in the adiabatic limit, since macroscopic fields are screened therein (Faraday-cage effect). A few calculations of Born charges in metals are available;\cite{Dreyer22,Marchese23} their possible experimental manifestations---e.g. in the case of a binary metal---are presently under discussion.\cite{Hickox23}

\bigskip
I thank Francesco Mauri for bringing this problem to my attention.
Work supported by the Office of Naval Research (USA) Grant No. N00014-20-1-2847.

\vfill \vfill
%\section*{References}
%\bibliography{$HOME/inputs/huge_bib}
%\bibliography{$HOME/inputs/huge_bib,add_bib}
%\bibliographystyle{unsrt}

\end{document}